# Performance Analysis of Probabilistic Rebroadcasting in Grid FSR for MANET

Nithya Rekha S [1] and Chandrasekar C [2]

[1] Computer Science Department, Periyar University
Salem, Tamilnadu 636011, India.
rekhasiva24@gmail.com

[2] Computer Science Department, Periyar University
Salem, Tamilnadu 636011, India.
ccsekar@gmail.com

**Abstract**
Mobile Ad-hoc Network (MANET) is the self organizing collection of mobile nodes. The communication in MANET is done via a wireless media. Ad hoc wireless **networks** have massive commercial and military potential because of their mobility support. Due to demanding real time multimedia applications, Quality of Services (QoS) support in such infrastructure less networks have become essential. QoS routing in mobile Ad-Hoc networks is challenging due to rapid change in network topology. In this paper, we focused to reduce flooding performance of the Fisheye State Routing (FSR) protocol in Grid using ns-2 network simulator under different performance metrics scenario in respect to number of Nodes. For example, the connection establishment is costly in terms of time and resource where the network is mostly affected by connection request flooding. The proposed approach presents a way to reduce flooding in MANETs. Flooding is dictated by the propagation of connection-request packets from the source to its neighborhood nodes. The proposed architecture embarks on the concept of sharing neighborhood information. The proposed approach focuses on exposing its neighborhood peer to another node that is referred to as its friend-node, which had requested/forwarded connection request. If there is a high probability for the friend node to communicate through the exposed routes, this could improve the efficacy of bandwidth utilization by reducing flooding, as the routes have been acquired, without any broadcasts. Friendship between nodes is quantized based on empirical computations and heuristic algorithms. The nodes store the neighborhood information in their cache that is periodically verified for consistency. Simulation results show the performance of this proposed method.

## 1. Introduction

The principal objective of a routing protocol is efficient discovery and establishment of a route between the source and the destination so that there can be a timely and efficient delivery of information between them. A Locating Service is used to locate the receiver inside the network. It dynamically maps the logical address of the receiver to its current location in the network. Once the receiver is located, routing and forwarding algorithms are used to route the information through the MANET. The routing is done using one-hop transmission service provided by the enabling technologies to construct an end-to-end (reliable) delivery services, from sender to one or more receivers. A number of features are expected to be supported by the routing protocols which include parameters like minimal control & processing overhead, loop freedom & prevention, efficient dynamic topology establishment and maintenance, scalability, support for unidirectional links, security & reliability and support for Quality of Service.[4,5]

The rest of this paper is organized as follows: First of all, we make a brief survey on FSR in section II with my previous research work. In section III, proposed work in grid FSR to reduce flooding. Section IV presents the Results and Discussion. Section V presents the Simulation Results of performance evaluation of various Parameters and section VI concludes the paper.

## 2. Related work

In my previous research work,[1] the investigations was on the behavior of the Proactive Routing Protocol Fisheye State Routing (FSR) in the Grid by analysis of various parameters. The Performance metrics that are used to evaluate routing protocols are Packet Delivery Ratio (PDR), Network Control Overhead, Normalized Overhead, Throughput and Average End to End Delay. Experimental results reveal that FSR is more efficient in Grid FSR in all QOS constraints. FSR can be used in all Resource critical environments. Grid Fisheye state routing (GFSR) consumes less bandwidth by restricting the propagation of routing control messages in paths formed by alternating gateways and neighbor heads, and allowing the gateways to selectively include routing table entries in their control messages. PDR and Throughput are 100% efficient in Simulation Results in NS2.

2.1 FSR (Fisheye State Routing)

G. P. Mario, M. Gerla and T.W. Chen, (2000) et.al, proposed that the FSR is a descendant of GSR [2]. In [3], the authors introduce a novel proactive (FSR), the notion of multi-level fisheye scope to reduce routing update overhead in large networks. Nodes exchange link state entries with their

This Research work was supported in part by Basic Scientist Research (BSR) Non SAP Scheme, under grant reference no.F-4-1/2006(BSR)/11-142/2010(BSR) UGC XI Plan.

neighbors with a frequency which depends on distance to destination. From link state entries, nodes construct the topology map of the entire network and compute optimal routes. Simulation experiments show that FSR is simple, efficient and scalable routing solution in a mobile, ad hoc environment. Fig. 1 refers the fisheye scope with different hops.

The following are the advantages of FSR.
* Simplicity
* Usage of up-to-date shortest routes
* Robustness to host mobility
* Exchange Partial Routing Update with neighbors

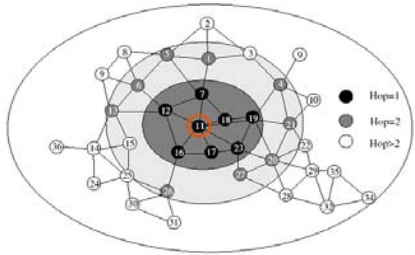

Fig. 1: Fisheye scope

## 3. Proposed work in Grid FSR to Reduce Flooding

The simulation results demonstrate the advantages of our approach. It can greatly reduce the redundant messages, thus saving much network bandwidth and energy. It can also enhance the reliability of broadcasting. It can be used in static or mobile wireless networks to implement scalable broadcast or multicast communications. As a result, the proactive approach provides a better quality of service by this new approach of Probability of calculating the Intimacy factor with neighbor node and friend node.

### 3.1 Caching & Timeouts

In order for the destination node to know the location of the destination or the receiver they have to acquire the route through the process of flooding. Flooding involves broadcasting of a packet to all the nodes of the network requesting the route of the destination node [6]. The nodes either respond with a reply back to the sender if in case, the current receiving node of the packet is the destination, or otherwise they forward the packet to other nodes. The destination node responds to the sender with the connection acknowledgement. The path traced by the acknowledge packet is remembered in all the forwarding hosts as the route from sender to destination. On one hand though the process of flooding helps the sender to dynamically obtain the location of the destination and the route over which information could be transmitted, it unnecessarily augments the load on the network as all the nodes in the network participate in the process of flooding[5]. If flooding is carried out frequently then it may altogether lead to the instability of the network. The direct implication of this observation is that flooding should be kept as infrequent as possible. One standard way to reduce the flooding mechanism is to provide the nodes with a small cache where the routes could be stored for future reference. The continuously changing characteristic of the ad hoc network environment poses further a problem, when routes are stored in a cache. There is always a possibility for the destination nodes to move from their place to another or even switch off. The cache reflects a static value and therefore in order to keep the data in the cache consistent they must be updated as frequently as possible. To account for these the routes in cache must be updated and validated periodically. The direct implication of this is that broadcast will be done frequently, which is to be avoided at all costs.

A new parameter referred to as the timeout period[6] is introduced to alleviate the problems arising. The timeout period is maintained for every route of the destination stored in the cache.This parameter reflects the lifetime of a route. The moment the timeout value expires the route in cache is deleted. Though still broadcasts are made, the frequency of broadcasts is reduced. The value of the timeout period reflects the frequency at which flooding occurs and if it is chosen to be a large number there is still a possibility of the route to have become invalid before the expiry of the cache. Therefore, the timeout value has to be prudently chosen. A tradeoff has therefore to be struck between the consistency of information and the frequency of broadcasts made.

### 3.2 Neighborhood

The suggested protocol falls under the category of reactive protocols. The nodes in this approach obtain the routes only when demand arises. The nodes use the common flooding approach to acquire the routes. Though flooding is used at the initial phases, it is decreased gradually. The most common method of using a cache with a fixed timeout period for each route is also used. The nodes are equipped with a small cache to save the routes and a timeout value is chosen. The selection of the timeout value is done as in case of the conventional networks, keeping in mind the other constraints of the network. However, the variation in the approach comes from the fact that the expiry of the timeout period does not trigger an update. The routes of the destination in the cache are rather erased after the timeout period. The nodes may then have to use flooding again to regain the routes, but in order to avoid that routes are shared between the nodes based on some criteria.

The primary focus of the protocol is on sharing information about the neighborhood of a peer with yet another node in the network. The neighborhood of a peer here reflects the contacts of the node in question with that of other nodes in the network. In other words, the neighborhood reflects the entries of routes in the cache. This information regarding the contacts a node has with other nodes in the vicinity is stored in tables or any other suitable data structure that is compatible with the protocol being adopted. The sharing of neighborhood information is not a mandatory task rather it is done at the discretion of the nodes concerned. The given approach intends to minimize the flooding requests that are needed to acquire the same information in the absence of the sharing mechanism.

The sharing of information occurs mainly between the nodes that are in direct contact with one another. Though the same can be carried out between nodes that are connected through a series of a finite number of intermediate nodes, factors such as the power levels of the nodes could be questioned to decide whether such a sharing could improve the efficiency or not. Since the sharing of information is at the discretion of the nodes in contact, they can decide whether the process has to be carried out or not. The process affects only the nodes that are communicating and any series of finite set of intermediate nodes that connect the two nodes, if such a process is to be triggered between nodes connected by multi-hop links. The nodes can also take into account parameters such as the current load on the route connecting the nodes, the current load on the two nodes and the power levels required to carry out the operation before initiating the sharing of information.

### 3.3 Intimacy Factor

The process of sharing of neighborhood information occurs when the receiving node decides that the node that started the communication between the two is ready for accepting the information. This decision could be made based on a parameter called the intimacy factor. The intimacy factor reflects the level of trust between the two nodes that are communicating. A threshold level of intimacy factor could be defined called as IFTHRES, which could then be compared against the intimacy factor, calculated between two communicating nodes to determine, when exactly to commence the sharing of neighborhood information. If the intimacy factor calculated is greater than IFTHRES then the receiving node can make a request to the sender enquiring its acceptance to the information about the routes to nearby nodes. This request is optional and the receiver does so at its discretion.

After the receiver ensures that the node that initiated the communication is ready to receive the neighborhood information, it posts a request to the sender. The sender can accept or reject the request. It can take into account the load on the link, the load on it, and its power level before posting to the receiver its consent. This can ensure that the sharing of routes doesn't exhaust the limited resources available. After the transmission of the sender's consent to the request, the sharing of the information or routes begins. The receiver shares a percentage of its cache entries with its friend node, depending upon the power levels and other such criteria. The sender then comes to know of the locations good possibility for the sender to send messages to these of various destinations close to the receiver. There may be a destinations, in which case the flooding process required for acquiring the same, have been eliminated.

### 3.4 Designing Approach

In a MANET, the presented approach could be modeled in the following way.
Total number of nodes in the network = $T_n$
Total number of nodes in cache = $K_n$
Unknown nodes = $U_n$

The network is considered to have $T_n$ number of nodes. The initiator of communication or the sender is assumed to have knowledge of routes of certain number of nodes in the network. The sender is unaware of the route of the other nodes, of which some may be near the receiver, with which the sender is currently communicating. The receiver is assumed to have a similar knowledge of routes of various nodes in the network. The set of routes in the receiver's cache need not be disjoint in comparison with the contents of the cache in the source's node; although the greater, the dissimilarity in the contents of the cache would imply a greater efficiency in the working of the protocol.

Route Gain Ratio (RGR) = (contents of sender's cache) ~ (contents of receiver's cache)
$$RGR \propto \eta$$
Where $\eta$, is the efficiency of the protocol.

After the receiving the routes of the nodes in the neighborhood of the receiver, these are stored in the cache of the sender. The basic understanding is that, given that the sender has contacted the receiver, it has a good probability to communicate with the nodes nearby the receiver. Since the approach is reactive protocol oriented, new routes have to be acquired before the transmission of information to the other destinations. Calculating the probability that the sender communicates with any of the unknown nodes or nodes for which it does not have the location, a clear understanding of the working efficiency of the protocol can be obtained.

Number of nodes (given) : $T_n$
Probability that an unknown node is contacted by the sender : $P_u$
The approach will prove to be efficient only if the sender
can utilize the information obtained from the receiver
before it expires.
Time available for the sender for utilizing the routes : $T_{out}$
Assuming the average time spent per node as, Average time spent in communicating with a node: $T_{avg}$
Total number of calls possible before routes expire :
$T_{out} / T_{avg} = T_c$
Total number of unknown nodes: $U_n$ (nodes whose route are unknown to sender)
Probty. that an unknown node is contacted : $P_u$
$P_u = ( U_n \, C \, Tcalls ) / ( T_n \, C \, U_n )$

When $T_n$ is large, $P_n$ tends to be very small. The efficiency of the protocol increases only when the unknown nodes contacted for a subset of the nodes in the neighborhood of the receiver. In other words, the maximum efficiency is gained only when the unknown node contacted is one of those exposed by the receiver to the sender during the sharing of neighborhood information.

Let number of nodes exposed = $E_n$
Probty. that a node exposed is contacted : $P_e$

$P_e = (E_n \, C \, T_c ) / (U_n \, C \, E_n )$
Probability that the node contacted forms a subset of the nodes exposed: $P = P_c * P_e$
If the probability that the node contacted is from the set of nodes whose routes have been exposed by the receiver, then the protocols succeeds in eliminating the flooding requests which otherwise would have been required to contact the unknown nodes. Considering the MANET environment to consist of a large number of nodes n and the probability $P_u$ being small, Poisson distribution could used to model the situation as following.
Total number of nodes = $n = T_n$
Probability that an exposed node is communicated = P
Let x be the number of exposed nodes contacted by the sender. Then, $P = \lambda$.
The set of routes that are exposed are only valid until the timeout period, after which they are deleted from the cache. The quantity of maximum concern here is the number of exposed nodes that are contacted.
Probty. that x nodes are contacted = P(X = x)
$P(X = x) = (e^{-\lambda} \lambda x) / x!$
$P(X = x) = (e^{-np} \lambda x) / x!$
$P(X = x) = (e^{-n(P_c * P_e)} \lambda x) / x!$

Where $P_c = (U_c\ C\ T_c) / (T_c\ C\ U_n)$   $P_e = (E_n\ C\ T_c) / (U_n\ C\ E_n)$
Total exposed nodes contacted: $T_e = P*E_n$

The higher the value of $T_e$, the lesser the broadcasts required for getting the routes for the unknown nodes. The probability that no exposed node is contacted is given by
$P(X = 0)$
$P(X = 0) = e^{-n(Pc * Pe)}$
$P_c * P_e > 0$ and always a finite quantity
$P(X = 0) = e^{-n(Pc*Pe)} = e > 0$

### 3.5 Increasing the Probability

The probability of contacting an exposed node is therefore never zero. To improve the probability and decrease further the flooding process that are carried out, the value of P(X=x) must be closer to unity.[6,10] To increase the number of exposed nodes contacted there exists two possible approaches, one by improving the value of En and the other wherein P is increased. Boosting the value of $E_n$ is not under the control of the designer. $E_n$ signifies the number of exposed nodes and is directly dependent on the neighborhood of the receiver that exposes the routes of the nodes to the sender. The value of En depends on the topology of the network, the density of the network and the mobility of the nodes in the network.

Although $E_n$ is strictly not under the control of the network designer, the value of En can be enhanced considerably by increasing the number of nodes exposed. In general, the receiver might then be expected to expose routes of the direct contacts it has, to the sender. In order to escalate further the probability of contacting an exposed node, it can augment the sample space of the nodes exposed. In other words, it can expose more nodes. This involves the receiver exposing nodes that are connected to it even through multi-hop links. The different nodes can be exposed one by one based on priorities assigned to them according to the distance of the exposed node from the receiver. The sender may wish to stop the transaction at any time in the middle by issuing an "I'm satiated" message. The receiver on receiving the message stops sending the routes.

The second method of increasing the probability P to improve the value of $T_e$ proves to be more feasible. In order to amplify the value of P the number of nodes that can be contacted before the exposed routes become invalid, can be boosted. This implies that the timeout period should be increased. If timeout value is enhanced then it can have two impacts on the network. The first impact is one, which would lead to lesser number of flooding, due to less frequent updates and a higher value of probability of contacting an exposed node. The second would promote a chance for the data or the routes to be corrupted between the timeout periods. As a consequence of this, a tradeoff has to be struck between consistency of data and the reduction of flooding requests.

### 3.6 Cases

The total number of messages that are transferred between the sender and receiver depends on the amount of information shared between them. It also depends on the number of intermediate nodes that are present between the sender and the receiver. However, considering the fact that the sharing of information only affects the sender, receiver and the finite number of intermediate nodes, if any is present; it can be concluded that the number of messages processed and transferred would be less than in case of flooding. This relies on the fact that the latter process involves all the nodes of the MANET environment. Therefore, even if the sharing process is a slightly prolonged one the process does not have any impact on the other nodes of the network, which still remain free for communication.

Also in the suggested approach, the flooding requests are minimized to a hop count of one. The flooding is initially limited to the immediate circle of nodes remaining in the coverage region of the node that is broadcasting the flooding request. Of these nodes, if any has the route to the destination in its cache it posts a message to the sender of the broadcast request, replying that the destination is its friend. At this point, the sender might establish the connection with the responding node, rather than the destination nodes itself. This intermediate node then routes the messages sent to it, to the intended destination.

In the worst case, the immediate circle of nodes in the coverage region of the source node might not have the route to the destination. Under such circumstances, the source node can rebroadcast the message with a hop count that can be found using an algorithm to increase the depth of flooding exponentially. When the flooding is done again with a greater hop count than used in the previous broadcast, the request is posted to more number of nodes. The process is repeated with a more number of nodes covered during each time.

The connection to the intended destination is then broken into different connections that involve nodes in direct contacts. Therefore, the source sends the message to the node with which it maintains the direct contact and which knows the route to the destination. This node then contacts with the destination or with the other node in the set of nodes connecting the source with the destination. The focus of the sender of the information is then on passing the information packet only to the node in direct contact. The task of routing the packet to the intended destination is then vested with the intermediate node that receives the packet. This responsibility then shifts from intermediate node to another one, if multiple nodes are present between the link connecting the sender and the receiver, as in case of any other multi-hop link. The last node meets with the responsibility, when it receives the packet and transmits it to the intended receiver through a direct link.

### 3.7 Friend & Stranger Nodes

In general, when two nodes start communicating with each other the sender or the initiator of the communication is moved to the stranger node state with respect to the receiver. As the communication proceeds, the intimacy factor is augmented based on some well-defined method. After the intimacy factor crosses the threshold value, the stranger node moves to the friend node state again with respect to the receiving node. This transformation between the states indicates that the receiver now is starting to trust the sender and share some information regarding the routes of nodes in its vicinity. The change of state triggers the sharing of routes, which is initiated by the receiver at the end of the ongoing transactions. The speed of this state change is a very important parameter in the design of the protocol. The faster the change, the earlier the sender or the initiator obtains the neighborhood information. This also has the consequence of a malicious node being able to quickly get the location of various destinations and launch an attack on the network. After the state change, the receiver is identified as being ready to receive the request for sharing the information regarding nearby nodes. The nodes that are acquired from the receiver are stored in the cache with a timeout period. Like any ordinary route that is stored in the cache after the expiry of the timeout period as per the norms of the protocol the routes are cleared.

The method of shifting the state of a source node or the initiator of a communication, from stranger node to friend node could be based either on some empirical or heuristic algorithms.

Empirically this could be done by maintaining a track of the messages transmitted between the nodes concerned or calculating the time during which the communication persists. It should also be noted that when the time of communication is taken into account, the factor could affect the sharing process In fact, it could bring down the efficiency of the protocol as the time to make use of the routes acquired is reduced. A balance therefore must be found between the two parameters. On the other hand, if the factor is based on the messages transmitted, a counter must be maintained by the receiver to count the packets received. In the aforementioned situation, the counter value could be directly used as the intimacy factor or could be weighted by any suitable constant to give the intimacy factor values.

Let the number of packets transmitted by the stranger node to receiver by Pt.

Pt $\propto$ k* Intimacy Factor, where k is some constant.

There also remains a good chance for the routes exposed to be already known to the sender. Under such circumstance, if possible the sender tries to correct the information that is maintained in the cache of the receiver. The sender then posts a "Gratis Reply" to the receiver. This informs the receiver the route, which was declared corrupt, and the new route that has to replace the corrupted one. A comparison is therefore required at the sender's side when it's receiving the exposed nodes' routes to ensure that the routes are correct. If during the comparison process the sender or the friend node to the receiver, identifies a route that is already known to it but is different from the one exposed by the receiver, it has to be able to discriminate between the right and the faulty route. The faulty one need not always be a wrong route, but can be an old route for which a newer version exists. To facilitate the required comparison an extra argument is used, which is referred to as the "Earmark". The earmark is calculated by subtracting the time at which a route was discovered from a standard reference time used across the network. A mechanism can be used to either accept a standard reference or to communicate a chosen reference across nodes whichever proves feasible.

## 4. Results and discussion

The Network simulator 2 has been used to analyze the parametric performance of Fisheye State Routing Protocol (FSR) in Grid .The metric based analysis is shown in figure 2 to figure 13. We simulate flooding protocols using Network Simulator 2 [15].

Moreover, performance of flooding protocols using Grid FSR has reduced flooding with respect to nodes. The nodes are increased from 50, 75,100,125,150. Thus, the expectation that the efficient flooding scheme has improved the Grid FSR performance with various parameters.

4.1 Performance Metrics
- End-to-End Delay: A specific packet is transmitting from source to destination and calculates the difference between send times and received times. Delays due to route discovery, queuing, propagation and transfer time are included in the delay metric. Certainly Fig. 2 and Fig.3 shows decrease in delay as in Flooding is reduced in FSR within Grid scenario.
- Packet Deliver Ratio (PDR): The (PDR) is defined as the ratio between the amount of packets sent by the source and received by the destination. Fig.4 and Fig.5 explains that PDR is more than 80% efficient in Reduced Flooding than FSR. On all other nodes PDR is better for FSR due to it scope technique and thus reduced traffic overhead.
- Throughput: Throughput is the average rate of successful data packets received at destination. It is usually measured in bits per second (bit/s or bps), and sometimes in data packets per second. The result are shown in Fig.6 and Fig.7.
- Jitter: Jitter is the variation of the packet arrival time. In jitter calculation the variation in the packet arrival time is expected to be low. The delays between the different packets need to be low for better performance in ad-hoc networks. It becomes a matter of concern if it is more than the threshold value, which is different for data. The result are shown in Fig. 8 and Fig.9.
- Energy: Based on realistic simulation models, these protocols shows significant energy-conserving potential. Energy is reduced in flooding becomes paramount in constraining battery dimensions and replenishment needs. Although, many routing protocols that minimize the energy consumed for multi-hop packet delivery have been designed, most of them surprisingly rely on flooding. The results are shown in Fig.10 and Fig.11 as energy is reduced due to disseminating flooding.
- Control Overhead: The results shows that overhead is reduced in Fig.12 and Fig.13. Network Control overhead (NCO) [1] is used to show the efficiency of the MANET's routing protocol scheme. It is defined, as the ratio of the number of control messages (the number of routing packets, Address Resolution Protocol (ARP), and control packets e.g., RTS, CTS and ACK) propagated by each node throughout the network and the number of the data packets received by the destinations. The reductions of network control overhead at higher data rate are very significant. This is because the same amounts of routing and control message are needed to route CBR traffic at lower data rate as well as at higher data rate. In reduced flooding, the control overhead can be reduced substantially.

## 5. Simulation results

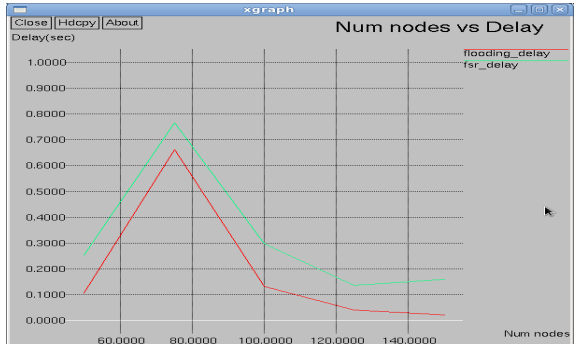
Fig. 2: Delay in Reduced Flooding with Grid FSR

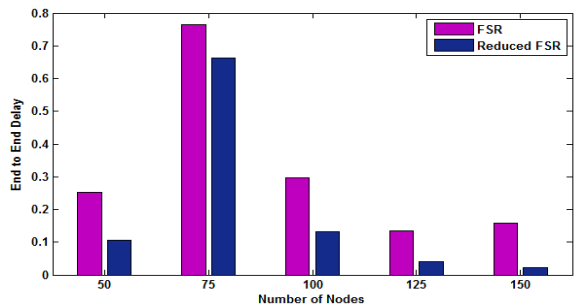
Fig. 3: End to End Delay V/s Nodes in Flooding

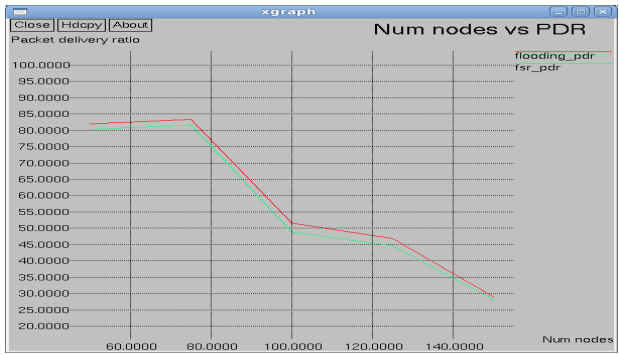
Fig. 4: PDR in Reduced Flooding with Grid FSR

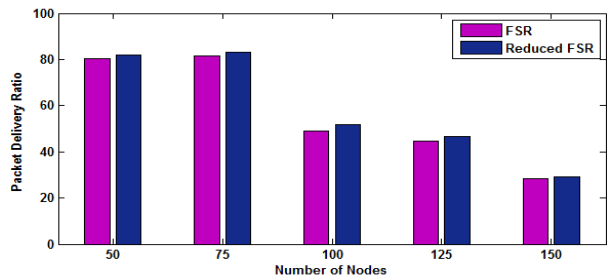
Fig. 5: Packet Delivery Ratio V/s Nodes in Flooding

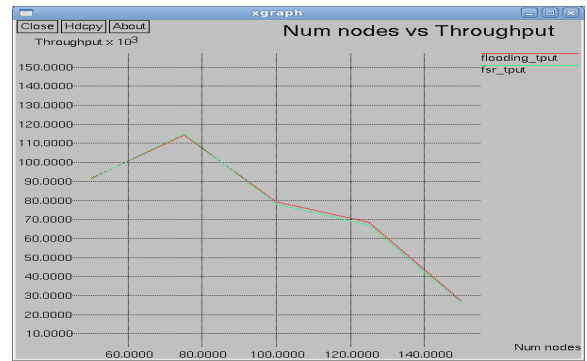
Fig. 6: Throughput in Reduced Flooding with Grid FSR

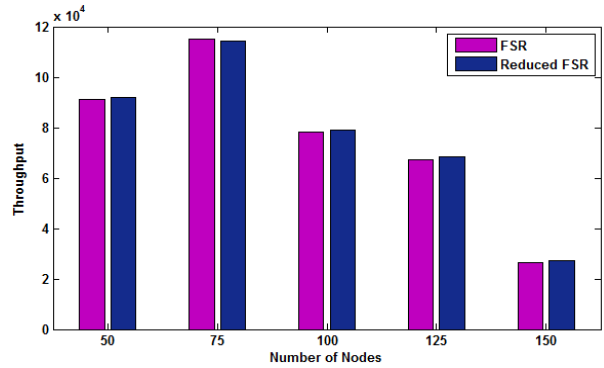
Fig.7: Throughput V/s Number of Nodes in Flooding

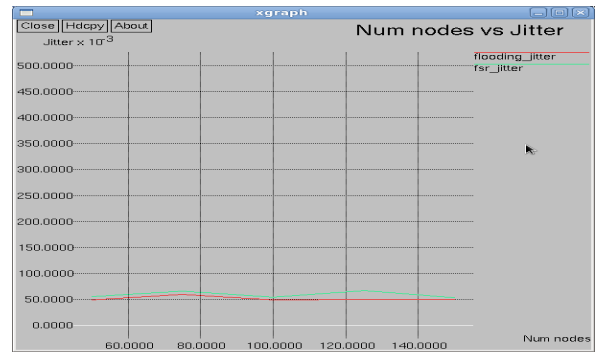
Fig. 8: Jitter in reduced Flooding with Grid FSR

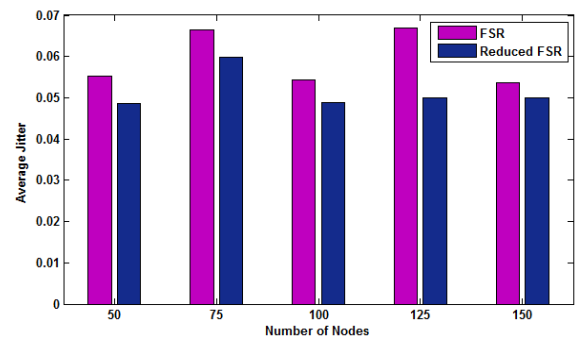
Fig. 9: Jitter V/s Nodes in Flooding

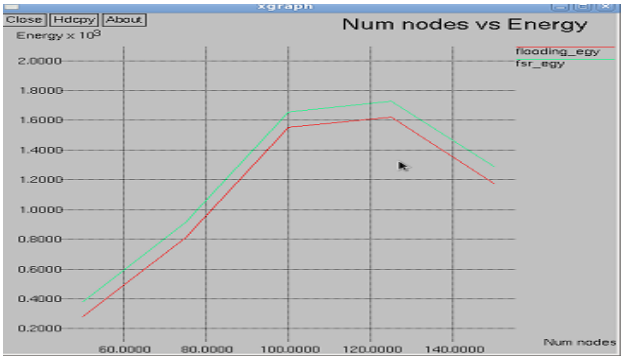
Fig. 10: Energy in Reduced Flooding with Grid FSR

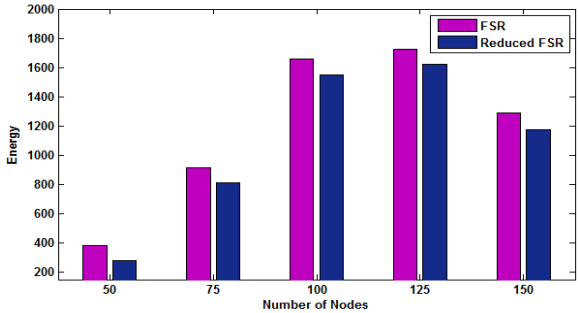
Fig.11: Energy V/s Number of Nodes in Flooding

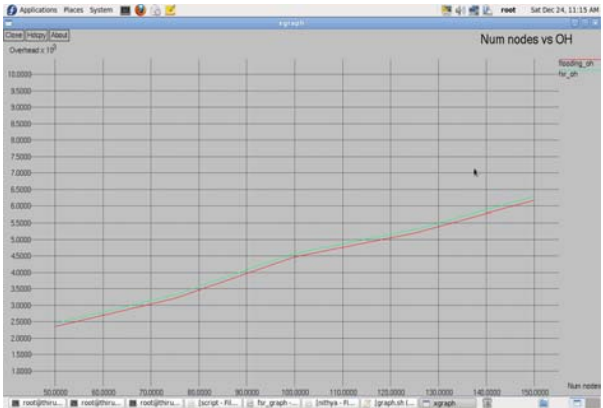
Fig. 12: Overhead in Reduced Flooding with Grid FSR

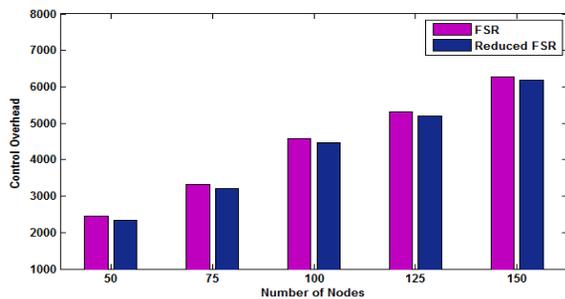
Fig. 13: Control Overhead V/s Nodes in Flooding

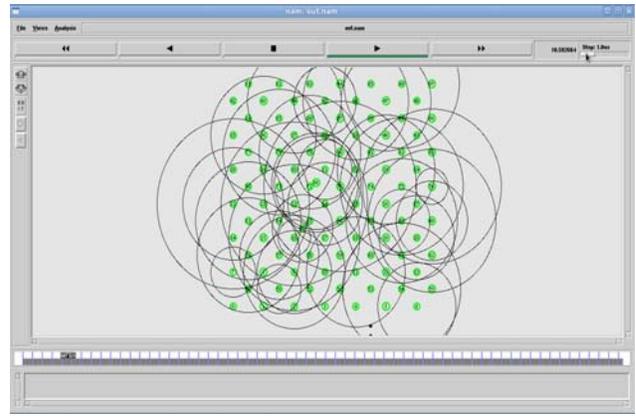
Fig. 14: Probability of Rebroadcasting in Nam Window

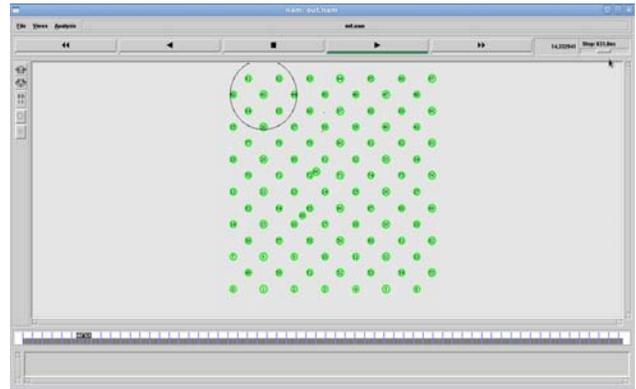
Fig. 15: Reduced Rebroadcasting in NAM Window

## 6. Conclusion and future work

Through a comparative study of efficient flooding, we have shown the following results. Figure 14 and 15 shows the output NAM Window of Rebroadcasting and reduced broadcasting.

Firstly, probability in Intimacy factor under certain circumstances, as the source node can rebroadcast the message with a hop count that can be found using an algorithm to increase the depth of flooding exponentially further to locate the destination. The protocol scales well for networks of all sizes. Secondly, The efficiency of the protocol has varying degrees of dependency with many parameters of the network, which includes the timeout period, the intimacy factor and others. Finally, the reduced flooding in routing traffic overhead and periodical propagation of link state information makes Grid FSR suitable for the high mobile dynamic changing network topology and thus the throughput is good with the high mobility of nodes, and therefore the average end-to-end delay is also very low. One of our future research works is to develop an efficient and optimized routing protocol with heavy mobility and routing overheads with different Mobility Models.

## Acknowledgment

The First Author extends her gratitude to UGC as this research work was supported by Basic Scientist Research (BSR) Non-SAP Scheme, under grant reference number, F-4-1/2006(BSR)/11-142/2010(BSR) UGC XI Plan.

**S.Nithya Rekha** is an IEEE member. She received her B.Sc Degree from Bharathiayar University in 1994, M.C.A. Degree in IGNOU and M.Phil Degrees from PRIST university, in 2006 and 2008, respectively. Her research interests include Mobile Computing, Rough set and Wireless networking. She has published three journals and presented papers in 3 International Conference and 8 National Conference. Her research work was supported by Basic Scientist Research (BSR) Non-SAP Scheme, under grant reference number,F-4-1/2006(BSR)/11-142/2010(BSR) UGC XI Plan as she is a Full-Time Ph.D. Research Scholar.

**Dr.C.Chandrasekar** is an IEEE member. He received his Ph.D degree from Periyar University. He has been working as Associate Professor, Department of Computer Science, Periyar University, Salem. His areas of interest include Wireless networking, Mobile Computing, Computer Communications and Networks. He is a research guide at various universities in India. He has published more than 40 technical papers at various National & International conferences and 50 journals.